\documentstyle[preprint,aps,prbbib]{revtex}                                                 
\begin{document}     
\begin{center}{\Large\bf Accelerating Universe, Cosmological Constant and Dark 
Energy}\end{center}
\begin{center}{Moshe Carmeli}\end{center}
\begin{center}{Department of Physics, Ben Gurion University, Beer Sheva 84105, 
Israel}\end{center}
\begin{center}{Email: carmelim@bgumail.bgu.ac.il}\end{center}
\begin{abstract}
Most of the calculations done to obtain the value of the cosmological constant
$\Lambda$ use methods of quantum gravity, a theory that has not been 
established as yet, and a variety of results are usually obtained. The numerical 
value of $\Lambda$ is then supposed to be inserted in the Einstein field 
equations, hence the evolution of the universe will depend on the calculated
value of $\Lambda$. Here we present a fundamental approach to the problem. The
theory presented here uses a Riemannian four-dimensional presentation of 
gravitation in which the coordinates are those of Hubble, i.e. distances and 
velocity rather than space and time. We solve these field equations and show
that there are three possibilities for the universe to expand but only the
accelerating universe is possible. We extract from the theory the cosmological 
constant and show that $\Lambda=2.036\times 10^{-35}s^{-2}$. This value of 
$\Lambda$ is in excellent agreement with the measurements obtained by the
{\it High-Z Supernova Team} and the {\it Supernova Cosmology Project}. Finally
it is shown that the three-dimensional space of the universe is flat, as the
Boomerang experiment shows.  
\end{abstract}
\section{Introduction}
The problem of the accelerating universe, the cosmological constant and the
vacuum energy associated with it is of extremely high interest these days. 
There are many questions related to these problems, especially with respect 
to the cosmological constant at the quantum level, all of which are related to 
quantum gravity. For example, why there exists the critical mass density and
why the cosmological constant has this or that value? Trying to answer these
questions and others were recently the subject of many publications [1-18]. 
Most of the calculations done to obtain the value of the cosmological constant 
$\Lambda$ use methods of quantum gravity. But there is no such a theory as yet
and hence a variety of results for $\Lambda$ are obtained. The numerical
value of $\Lambda$ is subsequently supposed to be inserted in the Einstein
field equations, thus determining the evolution of the universe and the vacuum
energy associated with it.

In this paper we present a fundamentally different approach in which the 
cosmological constant is not included {\it a preiori} in the theory, but $\Lambda$ 
can be extracted from the theory without the use of any quantum gravity 
theory. The value of $\Lambda$ will emerge from the gravitational field 
equations. Our theory employs a Riemannian four-dimensional presentation 
of gravity in which the coordinates are those of Hubble, namely distances and
(cosmological) velocity instead of the traditional space and time. The
gravitational field equations so constracted are subsequently solved and  
different kinds of expansion are obtained. It is shown that the accelerating
universe is the only acceptable possibility. We then extract the value of 
$\Lambda$ from the theory and show that it is given by 
$\Lambda=2.036\times 10^{-35}s^{-2}$, which is in excellent agreement with
the recently obtained measurements by the {\it High-Z Supernova Team} and the 
{\it Supernova Cosmology Project} [19-25]. Furthermore we determine the kind of the 
three-dimensional spatial subspace of the four-dimensional curved space to be
flat, which is in agreement with the recent Boomerang experiment [26,27].

In Section 2 a brief review of previous work is given. Section 3 is devoted to
the accelerated universe, whereas in Section 4 it is shown how the Tolman 
metric can be considered as an expanding universe. In Section 5 the value of
the cosmological constant is calculated, and the spatial three-dimensional 
subspace is determined to be flat. Section 6 is devoted to the 
concluding remarks.
\section{Review of Previous Work}
As in classical general relativity we start our discussion in flat space which
will then be generalized to curved space. 

{\bf The Flat Space Case.}
The flat-space cosmological metric is given by 
$$ds^2=\tau^2dv^2-\left(dx^2+dy^2+dz^2\right).\eqno(2.1)$$
Here $\tau$ is Hubble's time, the inverse of Hubble's constant, as given by
Freedman's measurements in the limit of zero distances and thus zero gravity 
[28,29]. As such, $\tau$ is a constant, in fact a universal constant (its
numerical value is given in Section 5). Its role 
in cosmology theory resembles that of $c$, the speed of light in vacuum, in 
ordinary special relativity. The velocity $v$ is used here in the sense of 
cosmology, as in Hubble's law, and is usually not the time-derivative of the 
distance.

The universe expansion is obtained from the metric (2.1) as a null condition,
$ds=0$. Using spherical coordinates $r$, $\theta$, $\phi$ for the metric 
(2.1), and the fact that the universe is spherically symmetric ($d\theta=
d\phi=0$), the null condition then yields $dr/dv=\tau$, or upon integration
and using appropriate initial conditions, gives $r=\tau v$ or $v=H_0r$, i.e.
the Hubble law in the zero-gravity limit.        

Based on the metric (2.1) a cosmological special relativity 
(CSR) was developed [30-36] (for applications to inflation in the universe
see Refs. [37-39]). In this theory the receding velocities of galaxies and the
distances between them in the Hubble expansion are united into a 
four-dimensional pseudo-Euclidean manifold, similarly to space and time in
ordinary special relativity. The Hubble law is assumed and is written in an
invariant way that enables one to derive a four-dimensional transformation 
which is similar to the Lorentz transformation. The parameter in the new 
transformation is the ratio between the cosmic time to $\tau$ (in which the 
cosmic time is measured backward with respect to the present time). 
Accordingly, the new transformation relates physical quantities at different 
cosmic times in the limit of weak or negligible gravitation. 

The transformation between the four variables $x$, $y$, $z$, $v$ and $x'$, 
$y'$, $z'$, $v'$ (assuming $y'=y$ and $z'=z$) is given by
$$x'=\frac{x-tv}{\sqrt{1-t^2/{\tau}^2}}, \ \ \ \
v'=\frac{v-tx/{\tau}^2}{\sqrt{1-t^2/{\tau}^2}}, \ \ \ \ 
y'=y, \ z'=z.\eqno(2.2)$$   
Equations (2.2) are the cosmological transformation and very much resemble the
well-known Lorentz transformation. In CSR it is the relative cosmic time which
takes the role of the relative velocity in Einstein's special relativity. The
transformation (2.2) leaves invariant the Hubble time $\tau$, just as the
Lorentz transformation leaves invariant the speed of light in vacuum $c$. 

{\bf Generalization to Curved Space.}
A cosmological general theory of relativity, suitable for the large-scale 
structure of the universe, was subsequently developed [40-43]. In the 
framework of cosmological general relativity (CGR) gravitation is described
by a curved four-dimensional Riemannian spacevelocity. CGR incorporates the
Hubble constant $\tau$ at the outset. The Hubble law is assumed in CGR as a
fundamental law. CGR, in essence, extends Hubble's law so as to incorporate 
gravitation in it; it is actually a distribution theory that relates distances 
and velocities between galaxies. The theory involves only measured quantities
and it takes a picture of the Universe as it is at any moment. The following 
is a brief review of CGR as was originally given by the author in 1966 in Ref. 
40.

The foundations of any gravitational theory are based on the principles of
equivalence and general covariance [44]. These two principles lead immediately
to the realization that gravitation should be described by a four-dimensional
curved spacetime, in our theory spacevelocity, and that the field equations 
and the equations of motion should be written in a generally covariant form.
Hence these principles were adopted in CGR also. Use is made in a 
four-dimensional Riemannian manifold with a metric $g_{\mu\nu}$ and a line 
element $ds^2=g_{\mu\nu}dx^\mu dx^\nu$. The difference from Einstein's general
relativity is that our coordinates are: $x^0$ is a velocitylike coordinate 
(rather than a timelike coordinate), thus $x^0=\tau v$ where $\tau$ is the
Hubble time in the zero-gravity limit and $v$ the velocity. The coordinate 
$x^0=\tau v$ is the 
comparable to $x^0=ct$ where $c$ is the speed of light and $t$ is the time in
ordinary general relativity. The other three coordinates $x^k$, $k=1,2,3$, are
spacelike, just as in general relativity theory.

An immediate consequence of the above choice of coordinates is that the null
condition $ds=0$ describes the expansion of the universe in the curved 
spacevelocity (generalized Hubble's law with gravitation) as compared to the 
propagation of light in the curved spacetime in general relativity. This means
one solves the field equations (to be given in the sequel) for the metric 
tensor, then from the null condition $ds=0$ one obtains immedialety the 
dependence of the relative distances between the galaxies on their relative
velocities.

As usual in gravitational theries, one equates geometry to physics.The first 
is expressed by means of a combination of the Ricci tensor and the Ricci
scalar, and follows to be naturally either the Ricci trace-free tensor or the
Einstein tensor. The Ricci trace-free tensor does not fit gravitation, and
the Einstein tensor is a natural candidate. The physical part is expressed by
the energy-momentum tensor which now has a different physical meaning from 
that in Einstein's theory. More important, the coupling constant that relates 
geometry to physics is now also different. 

Accordingly the field equations are
$$G_{\mu\nu}=R_{\mu\nu}-\frac{1}{2}g_{\mu\nu}R=\kappa T_{\mu\nu},\eqno(2.3)$$
exactly as in Einstein's theory, with $\kappa$ given by
$\kappa=8\pi k/\tau^4$, (in general relativity it is given by $8\pi G/c^4$),
where $k$ is given by $k=G\tau^2/c^2$, with $G$ being Newton's gravitational
constant, and $\tau$ the Hubble constant time. When the equations of motion 
will be written in terms of velocity instead of time, the constant $k$ will
replace $G$. Using the above equations one then has $\kappa=8\pi G/c^2\tau^2$.

The energy-momentum tensor $T^{\mu\nu}$ is constructed, along the lines of
general relativity  theory, with the speed of light being replaced by the
Hubble constant time. If $\rho$ is the average mass density of the universe,
then it will be assumed that $T^{\mu\nu}=\rho u^\mu u^\nu,$ where $u^\mu$ is
the four-velocity.
In general relativity theory one takes $T_0^0=\rho$. In Newtonian gravity one
has the Poisson equation $\nabla^2\phi=4\pi G\rho$. At points where $\rho=0$
one solves the vacuum Einstein field equations and the Laplace equation 
$\nabla^2\phi=0$ in Newtonian gravity. In both theories a null (zero) solution
is allowed as a trivial case. In cosmology, however, there exists no situation
at which $\rho$ can be zero because the universe is filled with matter. In
order to be able to have zero on the right-hand side of Eq. (2.3) one takes 
$T_0^0$ not as equal to $\rho$, but to $\rho_{eff}=\rho-\rho_c$, where 
$\rho_c$ is the critical mass density, 
a {\it constant} given by $\rho_c=3/8\pi G\tau^2$, whose value is 
$\rho_c\approx 10^{-29}g/cm^3$, a few hydrogen atoms per cubic meter. 
Accordingly one takes
$$T^{\mu\nu}=\rho_{eff}u^\mu u^\nu; \vspace{5mm}\rho_{eff}=\rho-\rho_c
\eqno(2.4)$$
for the energy-momentum tensor.

In the next sections we apply CGR to obtain the accelerating expanding 
universe and related subjects.
\section{The accelerating universe}
In previous works [40-43] it was shown that CGR predicts in a natural way that
the universe is accelerating. In this section we show that result in still 
another way. To this end we solve the gravitational field equations (2.3) with 
an energy-momentum tensor that includes pressure,
$$T^{\mu\nu}=\left[\rho_{eff}+p\right]u^\mu u^\nu-pg^{\mu\nu},
\eqno(3.1)$$
where $p$ is the pressure
The gravitational field that is sought is assumed to be both static and 
spherically symmetric and is therefore given by 
$$ds^2=e^\nu\tau^2dv^2-e^\lambda dr^2-r^2\left(d\theta^2+\sin^2\theta d\phi^2
\right)\eqno(3.2)$$
with $\nu$, $\lambda$ being functions of $r$ alone, and $u^0=u_0^{-1}=\left(
g_{00}\right)^{-1/2}$; other components of $u^\alpha$ are zero. As has been 
described in the last section, the universe expansion is obtained by the null
requirement, $ds=0$. Since, moreover, the universe expands in a spherically 
symmetric way, one also has $d\theta=d\phi=0$. As a result, Eq. (3.2) reduces
to 
$$e^\nu\tau^2dv^2-e^\lambda dr^2=0,\eqno(3.3)$$
which yields for the universe expansion the very simple formula
$$\frac{dr}{dv}=\tau e^{\left(\nu-\lambda\right)/2}.\eqno(3.4)$$

The nonvanishing components of the mixed Einstein tensor $G_\mu^\nu$ gives the
following for the gravitational field equations: 
$$G_0^0=-e^{-\lambda}\left(\frac{1}{r^2}-\frac{\lambda'}{r}\right)+
\frac{1}{r^2}=\kappa T_0^0,\eqno(3.5a)$$
$$G_0^1=-e^{-\lambda}\frac{\dot{\lambda}}{r}=\kappa T_0^1,\eqno(3.5b)$$
$$G_1^1=-e^{-\lambda}\left(\frac{\nu'}{r}+\frac{1}{r^2}\right)+
\frac{1}{r^2}=\kappa T_1^1,\eqno(3.5c)$$
$$G_2^2=-\frac{1}{2}e^{-\lambda}\left(\nu''+\frac{\nu'^2}{2}+
\frac{\nu'-\lambda'}{r}-\frac{\nu'\lambda'}{2}\right)+\frac{1}{2}e^{-\nu}
\left(\ddot{\lambda}+\frac{\dot{\lambda}^2}{2}-\frac{\dot{\nu}\dot{\lambda}}{2}
\right)=\kappa T_2^2,\eqno(3.5d)$$
$$G_3^3=G_2^2=\kappa T_3^3.\eqno(3.5e)$$
All other components of the Einstein tensor vanish identically, a prime 
denotes differentiation with respect to $r$, and a dot denotes differentiation
with respect to $v$.

The above equations then yield
$$e^{-\lambda}\left(\frac{1}{r^2}-\frac{\lambda'}{r}\right)-\frac{1}{r^2}=
-\kappa\rho_{eff},\eqno(3.6)$$
$$e^{-\lambda}\left(\frac{1}{r^2}+\frac{\nu'}{r}\right)-\frac{1}{r^2}=
\kappa p,\eqno(3.7)$$
$$e^{-\lambda}\left(\nu''+\frac{1}{2}\nu'^2+\frac{\nu'-\lambda'}{r}-
\frac{1}{2}\nu'\lambda'\right)=\kappa p.\eqno(3.8)$$
The conservation law $\nabla_\nu T^{\mu\nu}=0$ yields
$$p'=-\frac{1}{2}\nu'\left(p+\rho_{eff}\right).\eqno(3.9)$$
Equation (3.9) is not independent of Eqs. (3.6)-(3.8) since it is a 
consequence of the contracted Bianchi identities. One therefore has three
equations for the four unknown functions $\nu$, $\lambda$, $\rho$, $p$. One 
assumes a functional dependence of $\rho$ on $r$, calculate $\nu$, $\lambda$
from this knowledge, and finally calculate $p$.

The solution of Eq. (3.6) is given by
$$e^{-\lambda}=1-\frac{\kappa}{4\pi}\frac{m\left(r\right)}{r},\eqno(3.10)$$
where 
$$m\left(r\right)=4\pi\int_0^r\rho_{eff}\left(r'\right)r'^2dr'\eqno(3.11)$$
is the mass of the fluid contained in a ball of radius $r$. The solution given
by Eq. (3.10) is chosen so that $g_{\mu\nu}$ is regular at $r=0$ and goes to
the Schwarzschild form
$$e^{-\lambda}=1-\frac{r_s}{r},\eqno(3.12)$$
where $r_s=2Gm$ (divided by $c^2$) and $m=m\left(r_0\right)$, if 
$\rho_{eff}\left(r\right)=0$ for $r>r_0$.

We now assume that $\rho$ is a constant for $r\leq r_0$. We then obtain from
Eqs. (3.9), (3.6), (3.7) and (3.10) the following:
$$e^{-\lambda}=1-\frac{r^2}{R^2},\eqno(3.13)$$ 
$$e^{\nu/2}=A-B\left(1-\frac{r^2}{R^2}\right)^{1/2},\eqno(3.14)$$
$$p=\frac{1}{\kappa R^2}\left[
\frac{3B\left(1-\frac{r^2}{R^2}\right)^{1/2}-A}{A-B\left(1-\frac{r^2}{R^2}\right)^{1/2}}
\right],\eqno(3.15)$$
where $A$ and $B$ are constants, and
$$R^2=\frac{3}{\kappa\rho_{eff}}.\eqno(3.16)$$ 
The constants $A$ and $B$ can be fixed by the requirements that $p=0$ and 
$e^\nu$ join smoothly the Schwarzschild field on the surface of the sphere. 
One obtains
$$A=\frac{3}{2}\left(1-\frac{r_0^2}{R^2}\right)^{1/2},\vspace{5mm} 
B=\frac{1}{2},\eqno(3.17)$$
$$e^{\nu/2}=\frac{3}{2}\left(1-\frac{r_0^2}{R^2}\right)^{1/2}-
\frac{1}{2}\left(1-\frac{r^2}{R^2}\right)^{1/2},\eqno(3.18)$$
$$p=\rho\left[\frac{\left(1-\frac{r^2}{R^2}\right)^{1/2}-
\left(1-\frac{r_0^2}{R^2}\right)^{1/2}}{3\left(1-\frac{r_0^2}{R^2}\right)^{1/2}-
\left(1-\frac{r^2}{R^2}\right)^{1/2}}\right],\eqno(3.19)$$
with the condition that $r_0^2<R^2$. If one assumes that pressure inside the
fluid is everywhere finite, one obtains from Eq. (3.19) the more restrictive
condition
$$r_0^2<\frac{8}{9}R^2.\eqno(3.20)$$

The spacetime-coordinate version of the solutions presented above are due to 
K. Schwarzschild [45].

Using the above results in the equation for the universe expansion (3.4) we
obtain
$$\frac{dr}{dv}=\tau\left[A\left(1-\frac{r^2}{R^2}\right)^{1/2}-
B\left(1-\frac{r^2}{R^2}\right)\right].\eqno(3.21)$$
We now confine ourselves to the linear approxximation, getting
$$\frac{dr}{dv}=\tau\left(\tilde{A}+\tilde{B}\frac{r^2}{R^2}\right),
\eqno(3.22)$$
where $\tilde{A}=A-B$ and $\tilde{B}=B-A/2$, or
$$\frac{dv}{dr}=\tau^{-1}\left(\tilde{A}-\tilde{B}\frac{r^2}{R^2}\right).
\eqno(3.23)$$
A simplification is obtained if we also confine to the linear approximation of
$A$ and $B$, hence $A=3/2$, $B=1/2$, thus $\tilde{A}=1$, $\tilde{B}=-1/4$.
Using now the standard notation $\Omega=\rho/\rho_c$, we obtain
$$\frac{dr}{dv}=\tau\left[1+\frac{\left(1-\Omega\right)r^2}
{4c^2\tau^2}\right]\eqno(3.24)$$
for the equation of the expansion of the universe. Equation (3.24), except for 
the factor 4, is identical to Eq. (15) of Ref. 40 and Eq. (5.10) of Ref. 41.

The second term in the square bracket in the above equation represents the
deviation from the standard Hubble law due to gravity. For without that term,
Eq. (3.24) reduces to $dr/dv=\tau$, thus $r=\tau v+const$. The constant can be 
taken zero if one assumes, as usual, that at $r=0$ the velocity should also
vanish. Thus $r=\tau v$, or $v=H_0r$ (since $H_0=1/\tau$). Accordingly, the 
equation of motion (3.24) describes the expansion of the universe when 
$\Omega=1$, namely when $\rho=\rho_c$, the equation coincides with the
standard Hubble law.

The equation of motion (3.24) can easily be integrated exactly by the
substitions
$$\sin\chi =\left(\Omega-1\right)^{1/2}r/2c\tau;\Omega>1,\eqno(3.25a)$$
$$\sinh\chi =\left(1-\Omega\right)^{1/2}r/2c\tau;\Omega<1.\eqno(3.25b)$$
One then obtains, using Eqs. (3.24) and (3.25),
$$dv=cd\chi/\left(\Omega-1\right)^{1/2}\cos\chi;\Omega>1,\eqno(3.26a)$$
$$dv=cd\chi/\left(1-\Omega\right)^{1/2}\cosh\chi;\Omega<1.\eqno(3.26b)$$

We give below the exact solutions for the expansion of the universe for each 
of the case, $\Omega>1$ and $\Omega<1$. As will be seen, the case of 
$\Omega=1$ can be obtained at the limit $\Omega\rightarrow 1$ from both cases.

{\bf The case $\Omega>1$.}
From Eq. (3.26a) we have
$$\int dv=\frac{c}{\sqrt{\Omega-1}}\int\frac{d\chi}{\cos\chi},\eqno(3.27)$$
where $\sin\chi=r/a$, and $a=c\tau\sqrt{\Omega-1}$. A simple calculation gives
[47]
$$\int\frac{d\chi}{\cos\chi}=\ln\left|\frac{1+\sin\chi}{\cos\chi}\right|.
\eqno(3.28)$$
A straightforward calculation then gives
$$v=\frac{a}{2\tau}\ln\left|\frac{1+r/a}{1-r/a}\right|.\eqno(3.29)$$
As is seen, when $r\rightarrow 0$ then $v\rightarrow 0$ and using the
L'Hospital lemma, $v\rightarrow r/\tau$ as $a\rightarrow 0$ (and thus
$\Omega\rightarrow 1$).

{\bf The case $\Omega<1$.} 
From Eq. (3.26b) we now have
$$\int dv=\frac{c}{\sqrt{1-\Omega}}\int\frac{d\chi}{\cosh\chi},\eqno(3.30)$$ 
where $\sinh\chi=r/b$, and $b=c\tau\sqrt{1-\Omega}$. A straightforward 
calculation then gives [47]
$$\int\frac{d\chi}{\cosh\chi}=\arctan e^\chi.\eqno(3.31)$$  
We then obtain
$$\cosh\chi=\left(1+r^2/b^2\right)^{1/2},\eqno(3.32)$$
$$e^\chi=\sinh\chi+\cosh\chi=r/b+\left(1+r^2/b^2\right)^{1/2}.\eqno(3.33)$$
Equations (3.30) and (3.31) now give
$$v=\frac{2c}{\sqrt{1-\Omega}}\arctan e^\chi +K,\eqno(3.34)$$
where $K$ is an integration constant which is determined by the requirement
that at $r=0$ then $v$ should be zero. We obtain
$$K=-\pi c/2\sqrt{1-\Omega},\eqno(3.35)$$
and thus
$$v=\frac{2c}{\sqrt{1-\Omega}}\left(\arctan e^\chi-\frac{\pi}{4}\right).
\eqno(3.36)$$
A straightforward calculation then gives
$$v=\frac{b}{\tau}\left\{2\arctan\left(\frac{r}{b}+\sqrt{1+\frac{r^2}{b^2}}
\right)-\frac{\pi}{2}\right\}.\eqno(3.37)$$
As for the case $\Omega>1$ one finds that $v\rightarrow 0$ when $r\rightarrow 
0$, and again, using L'Hospital lemma, $r=\tau v$ when $b\rightarrow 0$ 
(and thus $\Omega\rightarrow 1$).

{\bf Physical meaning.}
To see the physical meaning of these solutions, however, one does not need the
exact solutions. Rather, it is enough to write down the solutions in the
lowest approximation in $\tau^{-1}$. One obtains, by differentiating Eq. 
(3.24) with respect to $v$, for $\Omega>1$,
$$d^2r/dv^2=-kr;\vspace{5mm}k=\left(\Omega-1\right)/c^2,\eqno(3.38)$$
the solution of which is 
$$r\left(v\right)=A\sin\alpha\frac{v}{c}+B\cos\alpha\frac{v}{c},\eqno(3.39)$$
where $\alpha^2=\Omega-1$ and $A$ and $B$ are constants. The latter can be
determined by the initial condition $r\left(0\right)=0=B$ and $dr\left(0
\right)/dv=\tau=A\alpha/c$, thus
$$r\left(v\right)=\frac{c\tau}{\alpha}\sin\alpha\frac{v}{c}.\eqno(3.40)$$
This is obviously a closed universe, and presents a decelerating expansion.

For $\Omega<1$ we have
$$d^2r/dv^2=\left(1-\Omega\right)r/c^2,\eqno(3.41)$$
whose solution, using the same initial conditions, is
$$r\left(v\right)=\frac{c\tau}{\beta}\sinh\beta\frac{v}{c},\eqno(3.42)$$
where $\beta^2=1-\Omega$. This is now an open accelerating universe.

For $\Omega=1$ we have, of course, $r=\tau v$.

We finally determine which of the three cases of expansion is the one at 
present epoch of time. To this end we have to write the solutions (3.40) and 
(3.42) in ordinary Hubble's law form $v=H_0r$. Expanding Eqs. (3.40) and 
(3.42) into power series in $v/c$ and keeping terms up to the second order, we
obtain
$$r=\tau v\left(1-\alpha^2v^2/6c^2\right) \eqno(3.43a)$$
$$r=\tau v\left(1+\beta^2v^2/6c^2\right) \eqno(3.43b)$$
for $\Omega>1$ and $\Omega<1$, respectively. Using now the expressions for 
$\alpha$ and $\beta$, Eqs. (3.43) then reduce into the single equation
$$r=\tau v\left[1+\left(1-\Omega\right)v^2/6c^2\right].\eqno(3.44)$$
Inverting now this equation by writing it as $v=H_0r$, we obtain in the lowest
approximation
$$H_0=h\left[1-\left(1-\Omega\right)v^2/6c^2\right],\eqno(3.45)$$
where $h=\tau^{-1}$. To the same approximation one also obtains
$$H_0=h\left[1-\left(1-\Omega\right)z^2/6\right]=h\left[1-\left(1-\Omega
\right)r^2/6c^2\tau^2\right],\eqno(3.46)$$
where $z$ is the redshift parameter.
As is seen, and it is confirmed by experiments, $H_0$ depends on the distance 
it is being measured; it has physical meaning only at the zero-distance limit,
namely when measured {\it locally}, in which case it becomes $h=1/\tau$.

It follows that the measured value of $H_0$ depends on the ``short"
and ``long" distance scales [48]. The farther the distance $H_0$ is being 
measured, the lower the value for $H_0$ is obtained. By Eq. (3.46) this is
possible only when $\Omega<1$, namely when the universe is accelerating. 

The possibility that the universe expansion is accelerating was first 
predicted using CGR by the author in 1996 [40] before the supernovae 
experiments results became known.

In the next section it is shown how the familiar Tolman metric can be looked
upon as an expanding universe.  
\section{Tolman's Metric as an Expanding Universe}
In the four-dimensional spacevelocity the Tolman metric is given by
$$ds^2=\tau^2dv^2-e^\mu dr^2-R^2\left(d\theta^2+\sin^2\theta d\phi^2\right),
\eqno(4.1)$$
where $\mu$ and $R$ are functions of $v$ and $r$ alone, and comoving 
coordinates $x^\mu=(x^0,x^1,x^2,x^3)=(\tau v,r,\theta,\phi)$ have been used. 
With the above choice of coordinates, the zero-component of the geodesic
equation becomes an identity, and since $r$, $\theta$ and $\phi$ are constants
along the geodesics, one has $dx^0=ds$ and therefore
$$u^\alpha=u_\alpha=\left(1,0,0,0\right).\eqno(4.2)$$
The metric (4.1) shows that the area of the sphere $r=constant$ is given by
$4\pi R^2$ and that $R$ should satisfy $R'=\partial R/\partial r>0$. The
possibility that $R'=0$ at a point $r_0$ should be excluded since it would
allow the lines $r=$constants at the neighboring points $r_0$ and $r_0+dr$ to
coincide at $r_0$, thus creating a caustic surface at which the comoving 
coordinates break down.

As has been shown in the previous sections the universe expands by the null
condition $ds=0$, and if the expansion is spherically symmetric one has
$d\theta=d\phi=0$. The metric (4.1) then yields
$$\tau^2 dv^2-e^\mu dr^2=0,\eqno(4.3)$$
thus
$$\frac{dr}{dv}=\tau e^{-\mu/2}.\eqno(4.4)$$
This is the differential equation that determines the universe expansion. In
the following we solve the gravitational field equations in order to find out
the function $\mu\left(r\right)$.

The gravitational field equations (2.3), written in the form
$$R_{\mu\nu}=\kappa\left(T_{\mu\nu}-\frac{1}{2}g_{\mu\nu}T\right),\eqno(4.5)$$
where 
$$T_{\mu\nu}=\rho_{eff}u_\mu u_\nu \eqno(4.6)$$
with $\rho_{eff}=\rho-\rho_c$ and $T=T_{\mu\nu}g^{\mu\nu}$, are now solved.
Using Eq. (4.2) one finds that the only nonvanishing components of $T_{\mu\nu}$ is
$T_{00}=\rho_{eff}$ and that $T=\rho_{eff}$.

The only nonvanishing components of the Ricci tensor are (dots and primes 
denote differentiation with respect to $v$ and $r$, respectively):
$$R_{00}=-\frac{1}{2}\ddot{\mu}-\frac{2}{R}\ddot{R}-\frac{1}{4}\dot{\mu}^2,
\eqno(4.7a)$$
$$R_{01}=\frac{1}{R}R'\dot{\mu}-\frac{2}{R}\dot{R}',\eqno(4.7b)$$
$$R_{11}=e^\mu\left(\frac{1}{2}\ddot{\mu}+\frac{1}{4}\dot{\mu}^2+\frac{1}{R}
\dot{\mu}\dot{R}\right)+\frac{1}{R}\left(\mu'R'-2R''\right),\eqno(4.7c)$$
$$R_{22}=R\ddot{R}+\frac{1}{2}R\dot{R}\dot{\mu}+\dot{R}^2+1-e^{-\mu}\left(
RR''-\frac{1}{2}RR'\mu'+R'^2\right),\eqno(4.7d)$$
$$R_{33}=\sin^2\theta R_{22},\eqno(4.7e)$$
whereas the Ricci scalar is given by
$$R=2e^{-\mu}\left[\frac{2}{R}R''+\left(\frac{R'}{R}\right)^2-\frac{1}{R}R'
\mu'\right]-\frac{2}{R}\dot{R}\dot{\mu}
-2\left(\frac{\dot{R}}{R}\right)^2-\frac{2}{R^2}-\frac{4}{R}\ddot{R}-
\ddot{\mu}-\frac{1}{2}\dot{\mu}^2.\eqno(4.8)$$

The field equations obtained for the components 00, 01, 11, and 22 (the 33 
component contributes no new information) are given by
$$-\ddot{\mu}-\frac{4}{R}\ddot{R}-\frac{1}{2}\dot{\mu}^2=\kappa \rho_{eff}
\eqno(4.9)$$
$$2\dot{R}'-R'\dot{\mu}=0 \eqno(4.10)$$
$$\ddot{\mu}+\frac{1}{2}\dot{\mu}^2+\frac{2}{R}\dot{R}\dot{\mu}+e^{-\mu}\left(
\frac{2}{R}R'\mu'-\frac{4}{R}R''\right)=\kappa\rho_{eff} \eqno(4.11)$$
$$\frac{2}{R}\ddot{R}+2\left(\frac{\dot{R}}{R}\right)^2+\frac{1}{R}\dot{R}
\dot{\mu}+\frac{2}{R^2}+e^{-\mu}\left[\frac{1}{R}R'\mu'-2\left(\frac{R'}{R}
\right)^2-\frac{2}{R}R''\right]=\kappa\rho_{eff} \eqno(4.12)$$
It is convenient to eliminate the term with the second velocity derivative of
$\mu$ from the above equations. This can easily be done, and combinations of 
Eqs. (4.9)--(4.12) then give the following set of three independent field 
equations:
$$e^\mu\left(2R\ddot{R}+\dot{R}^2+1\right)-R'^2=0 \eqno(4.13)$$
$$2\dot{R}'-R'\dot{\mu}=0 \eqno(4.14)$$
$$e^{-\mu}\left[\frac{1}{R}R'\mu'-\left(\frac{R'}{R}\right)^2-\frac{2}{R}R''
\right]+\frac{1}{R}\dot{R}\dot{\mu}+\left(\frac{\dot{R}}{R}\right)^2+
\frac{1}{R^2}=\kappa\rho_{eff} \eqno(4.15)$$
other equations being trivial combinations of (4.13)--(4.15).

The solution of Eq. (4.14) satisfying the condition $R'>0$ is given by
$$e^\mu=\frac{R'^2}{1+f\left(r\right)},\eqno(4.16)$$
where $f\left(r\right)$ is an arbitrary function of the coordinate $r$ and 
satisfies the
condition $f\left(r\right)>-1$. Substituting (4.16) in the other two field
equations (4.13) and (4.15) then gives
$$2R\ddot{R}+\dot{R}^2-f=0 \eqno(4.17)$$
$$\frac{1}{RR'}\left(2\dot{R}\dot{R'}-f'\right)+\frac{1}{R^2}\left(\dot{R}^2-f
\right)=\kappa\rho_{eff},\eqno(4.18)$$
respectively.

The integration of these equations is now straightforward. From Eq. (4.17) we
obtain the first integral
$$\dot{R}^2=f\left(r\right)+\frac{F\left(r\right)}{R},\eqno(4.19)$$
where $F\left(r\right)$ is another arbitrary  function of $r$. Substituting 
now (4.19) in Eq. (4.18) gives
$$\frac{F'}{R^2R'}=\kappa\rho_{eff}.\eqno(4.20)$$
The two Eqs. (4.19) and (4.20) are now integrated for the case 
for which $f$ equals to zero, and Eq. (4.19) consequently reduces to 
$$\dot{R}^2=\frac{F\left(r\right)}{R}.\eqno(4.21)$$

The integration of Eq. (4.21) gives
$$R\left(v,r\right)=\left[R^{3/2}\left(r\right)\pm\frac{3}{2}F^{1/2}\left(r
\right)v\right]^{2/3},\eqno(4.22)$$
where
$$R\left(r\right)=R\left(0,r\right),\eqno(4.23)$$
namely, $R\left(v,r\right)$ at $v=0$. Differentiating Eq. (4.22) with respect 
to $r$ and using Eq. (4.20) we also obtain
$$R\left(v,r\right)=\left(\kappa\rho_{eff}\right)^{-2/3}\left[\frac{R^{1/2}
\left(r\right)R'\left(r\right)}{F'\left(r\right)}\pm\frac{v}{2F^{1/2}
\left(r\right)}\right]^{-2/3}.\eqno(4.24)$$
Finally, from Eq. (4.20) we obtain
$$\frac{\partial}{\partial v}\left(\rho_{eff}R^2R'\right)=0,\eqno(4.25)$$
and accordingly one has
$$\frac{dr}{dv}=\tau e^{-\mu/2}=\tau\left(R'\right)^{-1/2}.\eqno(4.26)$$
If the function $f$ is not zero, the integration of Eq. (4.19) then yields for
$f>0$,
$$\tau v=f^{-1}\left(fR^2+FR\right)^{1/2}-Ff^{-3/2}\mbox{\rm arcsinh}
\left(fR/F\right)^{1/2}+\Phi\left(r\right), \eqno(4.27a)$$
and for $f<0$,
$$\tau v=f^{-1}\left(fR^2+FR\right)^{1/2}-F\left(-f\right)^{-3/2}\arcsin
\left(-fR/F\right)^{1/2}+\Phi\left(r\right), \eqno(4.27b)$$
where $\Phi\left(r\right)$ is an arbitrary function of $r$. The solutions 
(4.27) were given in spacetime coordinates by Datt [49].
\section{Value of $\Lambda$: Theory versus Experiment}
The Einstein gravitational 
field equations  
with the added cosmological term are [45]:
$$R_{\mu\nu}-\frac{1}{2}g_{\mu\nu}R+\Lambda g_{\mu\nu}=\kappa T_{\mu\nu},
\eqno(5.1)$$
where $\Lambda$ is the cosmological constant, the value of which is supposed to
be determined by experiment. In Eq. (5.1) $R_{\mu\nu}$ and $R$ are the Ricci 
tensor and scalar, respectively, $\kappa=8\pi G$, where $G$ is Newton's constant
and the speed of light is taken as unity.

Recently the two groups (the {\it Supernovae Cosmology Project} and the {\it 
High-Z Supernova Team}) concluded that the expansion of the universe is 
accelerating [19-25]. 
Both teams obtained
$$\Omega_M\approx 0.3,\hspace{5mm} \Omega_\Lambda\approx 0.7,\eqno(5.2)$$
and ruled out the traditional ($\Omega_M$, $\Omega_\Lambda$)=(1, 0)
universe. Their value of the density parameter $\Omega_\Lambda$ corresponds to
a cosmological constant that is small but, nevertheless, nonzero and positive,
$$\Lambda\approx 10^{-52}\mbox{\rm m}^{-2}\approx 10^{-35}\mbox{\rm s}^{-2}.
\eqno(5.3)$$

In Sections 2 and 3 a four-dimensional cosmological theory was presented. The 
theory predicts that the universe accelerates and hence it is equivalent to 
having a positive value for a cosmological constant in it.  
In the framework of this theory the 
zero-zero component of the field equations (2.3) is written as
$$R_0^0-\frac{1}{2}\delta_0^0R=\kappa\rho_{eff}=\kappa\left(\rho-\rho_c
\right),\eqno(5.4)$$
where $\rho_c=3/\kappa\tau^2$ is the critical mass density
and $\tau$ is Hubble's time in the zero-gravity limit.

Comparing Eq. (5.4) with the zero-zero component of Eq. (5.1), one obtains the
expression for the cosmological constant,
$$\Lambda=\kappa\rho_c=3/\tau^2.\eqno(5.5)$$

To find out the numerical value of $\tau$ we use the relationship between
$h=\tau^{-1}$ and $H_0$ given by Eq. (3.46) (CR denote values according to
Cosmological Relativity): 
$$H_0=h\left[1-\left(1-\Omega_M^{CR}\right)z^2/6\right],\eqno(5.6)$$
where $z$ is the redshift and $\Omega_M^{CR}=\rho_M/\rho_c$ where 
$\rho_c=3h^2/8\pi G$. (Notice that $\rho_c$ is different from 
the standard $\rho_c$ defined with $H_0$.) The redshift parameter $z$ 
determines the distance at which $H_0$ is measured. We choose $z=1$ and take 
for 
$$\Omega_M^{CR}=0.245 \eqno(5.7)$$
(roughly corresponds to 0.3 in the standard theory), Eq. (5.6) then gives
$$H_0=0.874h.\eqno(5.8)$$
At the value $z=1$ the corresponding Hubble parameter $H_0$ according to the 
latest results from HST can be taken [28] as $H_0=72$km/s-Mpc, thus 
$h=(72/0.874)$km/s-Mpc or
$$h=82.380\mbox{\rm km/s-Mpc},\eqno(5.9)$$
and 
$$\tau=12.16\times 10^9\mbox{\rm years}.\eqno(5.10)$$

What is left is to find the value of $\Omega_\Lambda^{CR}$. We have 
$\Omega_\Lambda^{CR}=\rho_c^{ST}/\rho_c$, where $\rho_c^{ST}=3H_0^2/8\pi 
G$ and $\rho_c=3h^2/8\pi G$. Thus $\Omega_\Lambda^{CR}=(H_0/h)^2=0.874^2$,
or
$$\Omega_\Lambda^{CR}=0.764.\eqno(5.11)$$
As is seen from Eqs. (5.7) and (5.11) one has 
$$\Omega_M^{CR}+\Omega_\Lambda^{CR}=1.009\approx 1,\eqno(5.12)$$
which means the universe is flat.

As a final result we calculate the cosmological constant according to 
Eq. (5.5). One obtains
$$\Lambda=3/\tau^2=2.036\times 10^{-35}s^{-2}.\eqno(5.13)$$

Our results confirm those of the supernovae experiments and indicate on
existance of the dark energy as has recently received confirmation from the
Boomerang cosmic microwave background experiment [26,27], which showed that 
the universe is flat.
\section{Concluding remarks}
In this paper it has been shown that application of the relativistic theory in
spacevelocity to the problem of the expansion of the universe yields on 
accelerating expansion. Furthermore, the cosmological constant that was
extracted from the theory agrees with the experimental result. Finally, it has
also been shown that the three-dimensional spatial space of the universe is
flat, again in agreement with observations. 

\end{document}